\documentclass[aps,twocolumn,pra,superscriptaddress,amsmath,showpacs,tightenlines]{revtex4}
\usepackage{graphicx}
\usepackage{color}

\begin{document}
\title{Preserving universal resources for one-way quantum computing}

\author{Tetsufumi Tanamoto}
\affiliation{Corporate R \& D center, Toshiba Corporation,
Saiwai-ku, Kawasaki 212-8582, Japan}

\author{Daniel Becker}
\author{Vladimir M. Stojanovi\'c}
\author{Christoph Bruder}
\affiliation{Department of Physics, University of Basel,
  Klingelbergstrasse 82, CH-4056 Basel, Switzerland}

\date{\today}

\begin{abstract}
The common spin Hamiltonians such as the Ising, $XY$, or Heisenberg
model do not have eigenstates that are suitable resources for
measurement-based quantum computation.  Various highly-entangled
many-body states have been suggested as a universal resource for this
type of computation, however, it is not easy to preserve these states
in solid-state systems due to their short coherence times.  To solve
this problem, we propose a scheme for generating a Hamiltonian that
has a cluster state as ground state.  Our approach employs a series of
pulse sequences inspired by established NMR techniques and holds
promise for applications in many areas of quantum information
processing.
\end{abstract}

\maketitle

\section{Introduction}

Measurement-based quantum computation (MQC) is a new computing
paradigm~\cite{Briegel_review}. Of particular interest are universal
resources of one-way quantum computation, a MQC scheme that requires
only local measurements~\cite{Briegel}.  In the original scheme of
one-way quantum computing, one initially creates a many-qubit cluster
state by applying phase gates or equivalent gate operations which can
be realized using the Ising interaction between qubits.  Many
promising methods to generate cluster states using solid-state qubits
have been proposed~\cite{borhani_loss,tana,Kay}.  However, since these
states are not the ground states of spin Hamiltonians with typical
qubit-qubit interactions of Ising, $XY$, and Heisenberg
form~\cite{Nest}, preserving them against the time evolution
generated by these spin Hamiltonians remains a critical issue.

One of the established universal resources are two-dimensional (2D)
cluster states.  Another promising candidate is the
Affleck-Kennedy-Lieb-Tasaki (AKLT) state on the honeycomb
lattice~\cite{AKLT}, a resonance valence bond (RVB) type state which
is a special projected entangled pair state
(PEPS)~\cite{Cirac,Rudolph}.  Yet, the AKLT state requires non-trivial
Hamiltonians with spin greater than 1/2, which are not easy to realize
in solid-state systems.

In this paper, we present a new method for preserving initially
prepared cluster states.
Our approach relies on manipulating a two-body Hamiltonian using
pulse-sequence techniques developed in the nuclear magnetic resonance
(NMR) context~\cite{Ernst,dinerman+santos}.  We show that, 
starting from the Ising and $XY$ models, one can induce an effective
dynamics described by a stabilizer Hamiltonian~\cite{Briegel}
\begin{equation}
H_{\rm stab}=-\sum_i K_i\:,
\label{stabH}
\end{equation}
where 
$K_i =\sigma^{x}_{i} \bigotimes_{j\in {\rm nbhd}(i)} \sigma^{z}_{j}$ 
are the correlation operators and the direct product runs over all 
nearest neighbors of the lattice site 
$i$ ($\sigma^{x}_i$ and $\sigma^{z}_j$ are the Pauli matrices).  
Combined with cluster state generation methods~\cite{Briegel_review,tana}, 
our scheme facilitates stable one-way quantum computing.  

We assume the original Hamiltonian to be of the form $H=H_0+H_{\rm int}$ where 
\begin{equation}
H_0= \sum_i (\Omega_i \sigma^{x}_{i} +\varepsilon_i \sigma^{z}_{i})
\label{H0}
\end{equation}
is a single-qubit part and $H_{\rm int}$ the interaction part.  We
take $H_{\rm int}$ to be of Ising $H_{\rm
  Ising}=\sum_{i<j}J_{ij}\sigma_i^z \sigma_j^z$, $XY$ $H_{\rm
  XY}=\sum_{i<j}J_{ij}[XY]_{ij}$, and Heisenberg form $H_{\rm
  H}=\sum_{i<j}J_{ij}[XYZ]_{ij}$.  In this paper, $J_{ij}=J$ if $i$
and $j$ are nearest neighbors and $J_{ij}=0$ otherwise.  We use the
shorthands $[XY]_{ij}\equiv \sigma_i^x\sigma_j^x+\sigma_i^y\sigma_j^y$
and $[XYZ]_{ij}\equiv \sigma_i^x\sigma_j^x
+\sigma_i^y\sigma_j^y+\sigma_i^z\sigma_j^z$, and set $\hbar=1$.

Note that a single correlation operator can be obtained 
using a single-qubit Hamiltonian. For example in a one-dimensional
(1D) qubit array, 
$K_2=\sigma_2^x \sigma_1^z \sigma_3^z$ can be generated
by the time evolution operator 
$e^{i(\pi/2) (\sigma_1^z+ \sigma_2^x+ \sigma_3^z)}$.
However, it is not evident how to obtain a sum like $K_2+K_3$ from
the single-qubit Hamiltonian.

Most fabricated solid-state qubit systems are nano-devices, because a
smaller size makes them more robust to decoherence.  An example are
quantum dot systems where smaller dots have larger energy-level
spacing.  Since with diminishing size it becomes difficult to address
these devices individually, it is of interest to consider switching
on/off $H_0$ and $H_{\rm int}$ independently.  We will show that, by
using appropriate pulse sequences, this is possible even if we start
from an $always$-$on$ Hamiltonian~\cite{Heule}.

\section{Method for preserving desirable states}

\begin{figure}
\includegraphics[width=3cm,clip=true]{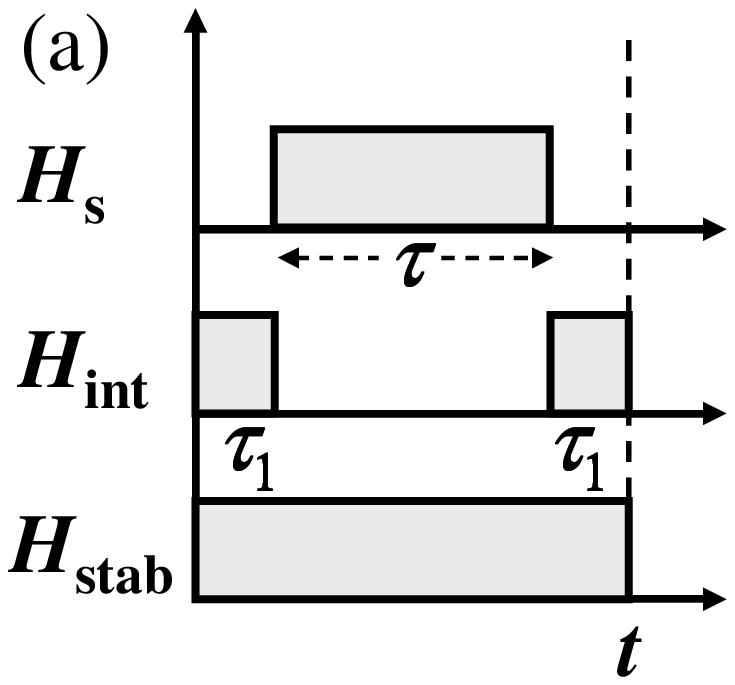}
\hspace{0.5cm}\includegraphics[width=4.5cm,clip=true]{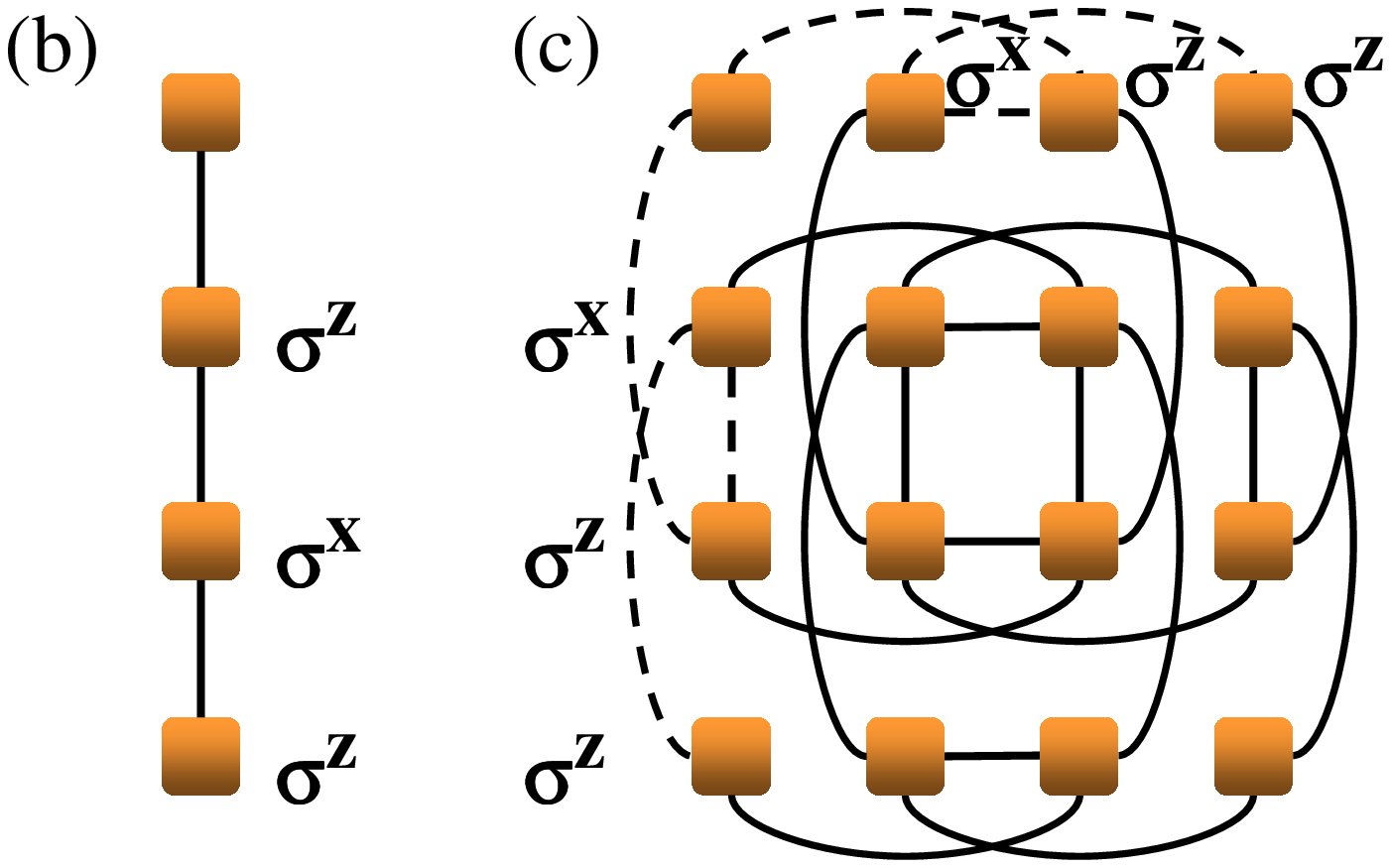}
\caption{(a) Switching on/off the parts $H_s$ and $H_{\mathrm{int}}$
  of the Hamiltonian $H=H_s+H_{\rm int}$ gives rise to an effective time
  evolution described by the stabilizer Hamiltonian
  $H_{\mathrm{stab}}$ [see Eq.~(\ref{main_result1})] whose ground state
  is a cluster state. 
(b) Graph representation of a one-dimensional 4-qubit (squares) cluster state
  stabilized by the Ising interaction.
(c) Two-dimensional $4\times 4$ twisted cluster state stabilized by
  the $XY$ interaction, a universal resource for one-way quantum computation. 
  The dashed lines and Pauli operators in each direction illustrate
  the twistedness of the state and the corresponding stabilizer
  Hamiltonian.} 
\label{Fig1}
\end{figure}

Desirable quantum states are preserved by a pulse sequence that
is familiar in the NMR context~\cite{Ernst}.  We assume that each
pulse is sufficiently strong such that interactions between qubits can
be neglected during the pulse sequences.  The time
evolution of the system is described by the density operator 
$\rho(t)$ whose time
dependence is given by $\rho(t) = \exp(-iHt)\rho(0)\exp(iHt)$ 
for time-independent $H$.  It is convenient to use
the following schematic notation for this evolution: $\rho(0)
\stackrel{t H}{\longrightarrow } \rho(t)$.  Then the process
\begin{equation}
\rho(0) \stackrel{\tau_1 H_{\rm int} }{\longrightarrow } \ \ 
\stackrel{ \tau H_s }{\longrightarrow } \ \ 
\stackrel{ -\tau_1 H_{\rm int}}{\longrightarrow } \rho(t)
\label{rho1}
\end{equation}
for $\tau_1=\pi/(4J)$ corresponds to $\rho(0) \stackrel{ \tau H_{\rm
    stab} }{\longrightarrow } \ \rho(t)$ where $t=\tau+\pi/(2J)$
and $\tau$ can be chosen arbitrarily.  
Note that at the physical time $t$ the state of the
system is obtained from the initial one by the time-evolution operator
\begin{equation}
e^{-i\tau H_{\mathrm{stab}}} = 
e^{-i\theta \sum_i \sigma^{z}_{i}\sigma^{z}_{i+1} } e^{-i\tau H_{\rm s}}
\left.e^{i\theta \sum_i \sigma^{z}_{i}\sigma^{z}_{i+1}}\right|_{\theta=\pi/4}\:.
\label{main_result1}
\end{equation}
Thus, as illustrated by Fig.~\ref{Fig1}(a), $H_{\mathrm{stab}}$ becomes
the effective system Hamiltonian. Its ground state is the originally
prepared cluster state, which is therefore preserved.

\subsection{Ising model}

We now show how to construct the 
stabilizer Hamiltonian using the relation
\begin{eqnarray}
& & e^{-i\theta \sigma^{z}_{1}\sigma^{z}_{2} } \sigma^{x}_{1} e^{i\theta \sigma^{z}_{1}\sigma^{z}_{2}} 
= \cos (2\theta) \sigma^{x}_{1} +\sin (2\theta) \sigma^{y}_{1} \sigma^{z}_{2},
\nonumber \\
& & e^{-i\theta \sigma^{z}_{1}\sigma^{z}_{2} } \sigma^{y}_{1} e^{i\theta \sigma^{z}_{1}\sigma^{z}_{2}} 
= \cos (2\theta) \sigma^{y}_{1} -\sin (2\theta) \sigma^{x}_{1} \sigma^{z}_{2}\:.
\label{Ising2}
\end{eqnarray}
An important consequence of these equations is that, for
$\theta=\pi/4$, we can increase the order of the Pauli-matrix terms as in 
$e^{-i\frac{\pi}{4}\sigma^{z}_{1}\sigma^{z}_{2}
} \sigma^{x}_{1} e^{i\frac{\pi}{4}\sigma^{z}_{1}\sigma^{z}_{2}} =
\sigma^{y}_{1} \sigma^{z}_{2} $ 
and 
$e^{-i\frac{\pi}{4}\sigma^{z}_{1}\sigma^{z}_{2} } \sigma^{y}_{1}
e^{i\frac{\pi}{4}\sigma^{z}_{1}\sigma^{z}_{2}} = -\sigma^{x}_{1}
\sigma^{z}_{2} $.  For a 1D $N$-qubit chain, the
starting single-qubit Hamiltonian is given by
\begin{equation}
H_{\rm s}= \Omega ( \sigma_1^y +\sum_{i=2}^{N-1} 
\sigma_i^x +\sigma_N^y)\:.
\label{hs}
\end{equation}
By switching the interaction, the 1D stabilizer Hamiltonian is
realized according to
\begin{equation}
H_{\mathrm{stab}} = 
e^{-i\theta \sum_i \sigma^{z}_{i}\sigma^{z}_{i+1}} H_{\rm s}
\left.e^{i\theta \sum_i \sigma^{z}_{i}\sigma^{z}_{i+1}}\right|_{\theta=\pi/4}\:,
\label{main_result2}
\end{equation}
the equivalent of Eq.~\eqref{main_result1} at the Hamiltonian level.
As an example, for $N=3$ qubits, starting from $H_{\rm s}=\Omega
(\sigma_1^y+\sigma_2^x+\sigma_3^y)$ we obtain $H_{\rm
  stab}=\Omega(-\sigma_1^x \sigma_2^z- \sigma_1^z
\sigma_2^x\sigma_3^z-\sigma_3^x \sigma_2^z)$.
If the system of $N$ qubits has periodic boundary conditions, we start
from the Hamiltonian $H_{\rm s}=\Omega \sum_{i=1}^{N} \sigma_i^x$.
Since Ising-type interaction terms commute and the time-evolution
operator $\exp(i(\pi/4) \sum_i \sigma^{z}_{i}\sigma^{z}_{i+1})$ in 
(\ref{main_result2}) factorizes, this process can be
straightforwardly extended to 2D and 3D qubit systems, thus realizing
the universal resource discussed in the introduction. Consequently,
for Ising interactions, we can construct the stabilizer Hamiltonian by
switching on $H_{\rm Ising}$ only once.

\subsection{$XY$ model}

Next, we show how to generate the stabilizer Hamiltonian
using the $XY$ interaction, assuming that $H_0$ and $H_{\mathrm{int}}$
can be switched on/off independently.

The stabilizer Hamiltonian is formed step by step by
bonding the nearest-neighbor operators.  This is because the $XY$
interactions do not commute, $[[XY]_{i-1,i}, [XY]_{i,i+1}]\neq
0$.  We start from
\begin{eqnarray}
e^{-i\theta [XY]_{12}} \sigma^{x}_{1} e^{i\theta [XY]_{12}} &=& 
\cos (2\theta) \sigma^{x}_{1} -\sin (2\theta) \sigma^{z}_{1} \sigma^{y}_{2},
\nonumber\\
e^{-i\theta [XY]_{12}} \sigma^{y}_{1} e^{i\theta [XY]_{12}} &=& 
\cos (2\theta) \sigma^{y}_{1} +\sin (2\theta)\sigma^{z}_{1}\sigma^{x}_{2}, 
\nonumber\\
e^{-i\theta [XY]_{12}} \sigma^{z}_{1} e^{i\theta [XY]_{12}} &=& 
\cos^2 (2\theta) \sigma^{z}_{1} +\sin^2 (2\theta) \sigma^{z}_{2}  \nonumber\\
&+&\frac{1}{2}\sin (4\theta)
[\sigma^{x}_{1}\sigma^{y}_{2}-\sigma^{y}_{1}\sigma^{x}_{2}]\:. 
\label{XY3} 
\end{eqnarray}
For $\theta=\pi/4$, these transformations increase the order
of the Pauli-matrix terms as 
$\sigma^{x}_{1} \rightarrow   -\sigma^{z}_{1} \sigma^{y}_{2}$ and 
$\sigma^{y}_{1} \rightarrow    \sigma^{z}_{1} \sigma^{x}_{2}$. 
For $\sigma_1^z$ one obtains $\sigma^{z}_{1} \rightarrow \sigma^{z}_{2}$.

We now show how to construct a 2D stabilizer Hamiltonian.
First we construct the 1D stabilizer Hamiltonian, starting from
\begin{equation}
H_{\rm s}=\Omega (-\sigma_1^x +\sigma_2^y -\sum_{i=3,N-2} 
\sigma_i^x +\sigma_{N-1}^y -\sigma_N^x)\:.
\end{equation}
In the specific case of six qubits in 1D, by applying 
Eq.~(\ref{XY3}) to $[XY]_{12}$, $[XY]_{34}$, and $[XY]_{56}$, we obtain:
\begin{eqnarray}
e^{-iS_1} H_{\rm s} e^{iS_1}
&=&\Omega ( \sigma^{z}_{1}\sigma^{y}_{2} +\sigma^{z}_{2}\sigma^{x}_{1} 
+ \sigma^{z}_{3}\sigma^{y}_{4} + \sigma^{z}_{4}\sigma^{y}_{3}
\nonumber \\
& +& \sigma^{z}_{5}\sigma^{y}_{6} + \sigma^{z}_{6}\sigma^{y}_{5}) \:, 
\label{S_1}
\end{eqnarray} 
where 
$S_1=\frac{\pi}{4} \sum_{l=1,3} [XY]_{\{2l-1,2l\}}$. 
Repeating this step with 
$S_2=\frac{\pi}{4} \sum_{l=1,2} [XY]_{\{2l,2l+1\}}$, 
we get the 1D stabilizer Hamiltonian
$H_{\rm 1 D} = e^{-iS_2} e^{-iS_1} H_{\rm s} e^{iS_1}e^{iS_2}$
that reads explicitly
\begin{eqnarray}
H_{\rm 1 D} &=& \Omega (
\sigma^{z}_{1}\sigma^{z}_{2}\sigma^{x}_{3}+\sigma^{z}_{3}\sigma^{x}_{1}
+ \sigma^{z}_{2}\sigma^{z}_{4}\sigma^{x}_{5}+
\sigma^{z}_{5}\sigma^{z}_{3}\sigma^{x}_{2}\nonumber \\ 
&+&
\sigma^{z}_{4}\sigma^{x}_{6} +\sigma^{z}_{6}\sigma^{z}_{5}\sigma^{x}_{4}) \:.
\end{eqnarray}
This Hamiltonian is twisted in the sense of \cite{tana}, i.e., the
site indices of the corresponding cluster state are obtained by the
permutation $(2,3)(4,5)\ldots(N-2,N-1)$ (cyclic notation), for a chain 
of $N$ qubits where $N$ is even, see Fig.~\ref{Fig1}(b,c).  

The next step in the construction of the 2D stabilizer Hamiltonian is
to construct a ladder Hamiltonian by bonding nearest-neighbor sites on
adjacent chains $a$ and $b$, in which all the bondings between qubits
$ia$ and $ib$ are carried out simultaneously:
\begin{eqnarray}
H_{\rm ladder}&=&\Omega(
-\sigma^{z}_{1a}\sigma^{z}_{2a}\sigma^{z}_{b3}\sigma^{y}_{3a}
-\sigma^{z}_{3a}\sigma^{z}_{1b}\sigma^{y}_{1a} 
\nonumber \\
&-&\sigma^{z}_{2a}\sigma^{z}_{4a}\sigma^{z}_{5b}\sigma^{y}_{5a}
-\sigma^{z}_{5a}\sigma^{z}_{3a}\sigma^{z}_{2b}\sigma^{y}_{2a}
-\sigma^{z}_{4a}\sigma^{y}_{6a}\sigma^{z}_{6b}
\nonumber \\
&-&\sigma^{z}_{6a}\sigma^{z}_{5a}\sigma^{z}_{4b}\sigma^{y}_{4a}) + 
(a \leftrightarrow b)\:.
\end{eqnarray}

A 2D stabilizer Hamiltonian is produced by connecting the above two
ladder Hamiltonians with the interaction between the two ladders.  For
example, when we prepare two ladders of length 4 such as in
Fig.~\ref{Fig1}(c) and connect them vertically, we obtain a $4\times
4$ stabilizer Hamiltonian.

\subsection{Heisenberg model}

For the Heisenberg interaction, we can construct only a two-qubit
stabilizer Hamiltonian (note that the same is true for the $XXZ$
interaction).  The basic relation is
\begin{eqnarray}
& & e^{-i\theta [XYZ]_{12}} \sigma^{y}_{1} e^{i\theta [XYZ]_{12}} = 
\cos^2 (2\theta) \sigma^{y}_{1}  \nonumber \\
& & +\sin^2 (2\theta) \sigma^{y}_{2} 
+\frac{1}{2} \sin (4\theta) (\sigma^{x}_{1}\sigma^{y}_{2}-
\sigma^{y}_{1}\sigma^{x}_{2})\:. 
\label{Heisenberg1}
\end{eqnarray}
For the Ising and $XY$ models, we can eliminate the single Pauli matrix
terms leaving the interaction terms [see Eqs.~(\ref{Ising2}) and
(\ref{XY3})]. 
However, in Eq.~(\ref{Heisenberg1}), if we set $\sin (2\theta)=0$ or $\cos
(2\theta)=0$, we also eliminate the
$\sigma^{x}_{1}\sigma^{y}_{2}-\sigma^{y}_{1}\sigma^{x}_{2}$ term.
This is because the Heisenberg interaction contains terms in all three
spatial directions~\cite{tana}. In the case of two qubits, we obtain
$H=\Omega(\sigma^z_1 \sigma^x_2 -\sigma^x_1 \sigma^z_2)$
from the initial Hamiltonian 
$H_s= \Omega (\sigma^y_1 -\sigma^y_2)$
by using Eq.~(\ref{Heisenberg1}) for $\theta=\pi/8$.  By applying a
$\pi$-rotation, we obtain the two-qubit stabilizer Hamiltonian
$H=\Omega(\sigma^z_1 \sigma^x_2 +\sigma^x_1 \sigma^z_2)$.

\section{Manipulation of always-on Hamiltonian}

The scheme discussed up to now relies on switching on/off the single-qubit
Hamiltonian $H_{0}$ [see Eq.~(\ref{H0})] 
and the Ising or $XY$ interaction part $H_{\rm int}$
separately. There is a number of schemes for switching on/off
interactions between qubits (see, e.g., \cite{averin_bruder,Nakamura,nori}). 
However, they make the system more complicated and require
additional overhead. 

Here, we solve this problem by demonstrating how to extract $H_0$ and
$H_{\rm int}$ by using appropriate pulse sequences.  We illustrate
the idea using the standard NMR Hamiltonian $H_{\rm nmr}=\sum_i
\varepsilon_i \sigma_i^z+ \sum_i J \sigma_i^z \sigma_{i+1}^z$ which has
the property that $[H_0, H_{\rm int}]=0$.  In this case, $H_0$ and
$H_{\rm int}$ can be switched on/off by using a simple pulse sequence.
The interaction part $H_{\rm Ising}$ can be extracted by using two 
sandwiched $\pi$-pulses such as 
$\exp(i\tau H_{\rm Ising})=e^{-i(\pi/2) \sum_j \sigma_j^y} e^{ i(\tau/2) 
H_{\rm nmr}} e^{ i(\pi/2) \sum_j \sigma_j^y} e^{ i(\tau/2) H_{\rm nmr}}$. 
On the other hand, two steps are required to obtain $H_0$. 
Let us consider a 1D qubit chain. By applying a $\pi$-pulse 
about the $x$-axis (denoted by $(\pi)_x$) to all the qubits on the even sites, 
we obtain 
$e^{ -i(\pi/2) \sum_{i} \sigma_{2i}^y} e^{ i(\tau/2) H_{\rm nmr}} 
e^{ i(\pi/2) \sum_{i} \sigma_{2i}^y} e^{ i(\tau/2) H_{\rm nmr}} 
=e^{i\tau \sum_i \Omega \sigma_{2i-1}^z} 
$. 
Similarly, we obtain $e^{i\tau \sum_i \Omega \sigma_{2i-1}^z}$ 
by applying  a $(\pi)_x$-pulse to all the qubits on the odd sites.
Combining these two processes yields $H_0$. 
This method is easily generalized to 2D or 3D qubit arrays.

If $[H_0,H_{\rm int}]\neq 0$, this NMR method cannot be used.
Even in this case, $H_0$ and $H_{\rm int}$ can be extracted separately. 
The idea follows from average Hamiltonian
theory  which is based on the Baker-Campbell-Hausdorff
(BCH) formula for the expansion of $e^Ae^B$ \cite{Ernst}.
A stroboscopic application of the Hamiltonian designed by 
a series of short pulses can reduce or eliminate unwanted terms, 
if $\Omega\tau, J\tau \ll 1$.
First we extract $H_{\rm int}$ by setting 
$A=h_0+h_1$ and $B= -h_0+h_1$ in the BCH formula, 
where $h_0\equiv H_0 \tau$ and $h_1\equiv H_{\mathrm{int}} \tau$. 
$B$ is realized by applying a $(\pi)_y$-pulse on every qubit.  
From the BCH formula, we obtain
\begin{equation}
e^Ae^B \approx \exp ( 2h_1 +[h_0, h_1]+\frac{1}{3} [h_0, [h_0,h_1]])\;.
\label{AB}
\end{equation} 
The exponent corresponds to a third-order expansion in $\Omega\tau$
for $\Omega \tau \ll 1$.  If we repeat this operation $n$ times like
$e^Ae^Be^Ae^B\cdots e^Ae^B=(e^Ae^B)^n$ such that $n\Omega\tau=\pi/4$,
the $k$-th term is of order $[\pi/(4n)]^k$. Therefore, $H_0$ is
canceled, and we obtain only $H_{\rm int}$ in this order.  When we
apply
\begin{equation}
e^Ae^B e^Be^A \approx \exp (4h_1 
-\frac{5}{3} [h_1, [h_0,h_1]]+\frac{1}{3} [h_0, [h_0,h_1]])\:,
\end{equation} 
we can eliminate the second term in Eq.~(\ref{AB}).  In the limit
$n\rightarrow \infty$ under the condition of $n\Omega\tau =\pi/4$,
$H_0$ is exactly eliminated. The extraction of $H_0$ can be achieved
analogously. 
Moreover, as shown in \cite{Waugh}, 
if the $k$th-order term is the first
nonvanishing correction, the decay rate $T_2$ of the qubit system
is enlarged according to $T_2' \propto T_2(k+1)! (T_2/t_c)^k$ as long
as $T_2>t_c$ where $t_c$ is the time required for each single step 
[$A$ and $B$ in Eq.~(\ref{AB})].

For the $XY$ model, we have to switch off subsets of $H_{\rm int}$
corresponding to $S_1$ and $S_2$, as discussed after Eq.~(\ref{S_1}).
This is equivalent to choosing $A$ and $B$ appropriately: e.g., for
the 1D chain, $A=H_{\mathrm{int}}=h_{1e}+h_{1o}$ and $B=h_{1e}-h_{1o}$
where 
$h_{1e}=J \tau([XY]_{23}+[XY]_{45}+[XY]_{67}+..)$ and
$h_{1o}=J \tau([XY]_{12}+[XY]_{34}+[XY]_{56}+..)$.  That means, $B$
is generated by applying $\pi$ pulses to qubits $2,3,6,7,10,11,...$.

Let us consider the 1D $XY$ model with $\varepsilon_i=0$ in Eq.~(\ref{H0}).  
The following operation can be used to obtain $H_0$.  
(i) Applying a
$(\pi)_x$-pulse to all the qubits on the even sites changes the sign
of $\sum_i \sigma_i^y\sigma_{i+1}^y$.  (ii) By further applying a
$(\pi)_y$-pulse to the same subset of qubits, we obtain $B/\tau
=\sum_i \Omega (\sigma_{2i-1}^x -\sigma_{2i}^x)-\sum_iJ[XY]_{i,i+1}$.
As a result, we obtain $\Omega\sum_i \sigma_{2i-1}^x$.  Repeating the
same operations with the qubits on the odd sites, we obtain $\Omega
\sum_i \sigma_{2i}^x$.  For the Ising model, the process (i) is not
required. For the Heisenberg model, the same procedure as in the case
of the $XY$ model does not eliminate the term $J\sum_i \sigma_i^z
\sigma_{i+1}^z$.  Thus, additional similar steps are required to eliminate
this term.  

\section{Robustness}

Since a practical realization of these pulse sequences will not be
free of imperfections, we now analyze the effect of pulse duration
errors $\delta$.  
A central quantity will be the cluster-state fidelity  
$F_{\rm st}(\tau)=
|\langle\Psi_{00...0}|U_\tau(\delta)|\Psi_{00...0}\rangle|^2$ where 
$U_\tau(\delta)=e^{-i\tau H_{\mathrm{stab}}(\pi/4+\delta/4)}$.
In the Ising case, $H_{\mathrm{stab}}$ is given by
Eq.~(\ref{main_result2}), and an analogous equation in the $XY$ case.

Let us consider the 1D Ising case.  For $\theta=\pi/4 +\delta/4$, the
first-order correction to Eq.~(\ref{main_result2}) reads
\begin{equation}
H'(\delta)=\frac{\delta}{2} \Omega\left[\sigma_1^y +\sigma_N^y 
+\sum_{1<i<N} \sigma_i^y(\sigma_{i-1}^z+\sigma_{i+1}^z)\right]\:.
\label{Hprime}
\end{equation}
The effect of these terms is calculated from perturbation theory using
the expressions for $\sigma_i^\alpha|\Psi_{00..0}\rangle$, where
$|\Psi_{00...0}\rangle$ is the initial cluster state,
{generated e.g. as proposed in} Ref.~\cite{tana},  and
$\alpha=x,y,z$ \cite{Aschauer}.  The lowest-order expression of the
cluster-state fidelity reads $F_{\rm st}(\tau) \approx \left[ 1+
  \Omega^2 \tau^2 \delta^2 (N-1)/2 \right]^{-2}$, and the correction
scales with $\delta^2$ which is a signature of the robustness of our
method.

The simplest and most powerful method to further reduce
the effect of pulse imperfections is the symmetrization of the pulse
sequence frequently used in NMR~\cite{Ernst}.
We first note that Eq.~(\ref{rho1}) is equivalent to
\begin{equation}
\rho(0) \stackrel{\tau_1 H_{\rm int}}{\longrightarrow } \ \ \stackrel{ \frac{\tau}{2} H_0 }{\longrightarrow } \ \ 
\stackrel{ -\tau_1 H_{\rm int} }{\longrightarrow }\  
\stackrel{ -\tau_1 H_{\rm int} }{\longrightarrow } \ \ \stackrel{ \frac{\tau}{2} H_0 }{\longrightarrow } \ \ 
\stackrel{ \tau_1 H_{\rm int} }{\longrightarrow } \rho(t)\:,
\label{rho2}
\end{equation}
where, as before, $\tau_1=\pi/(4J)$.  The second half of this pulse
sequence results in a perturbation term that has the opposite sign as
compared to Eq.~(\ref{Hprime}).  Applying Eq.~(\ref{AB}) leads to a
cancellation of the first-order perturbation term.  If the original
interval length $\tau$ is divided into an even number $n$ of subintervals,
$\tau=nt_c$, the perturbation term is replaced by $[H_{\mathrm{stab}},
  H'] i\tau\delta/(4n)$, and the fidelity $F_{\rm sym}^{n}$ is given
by
$F_{\rm sym}^n \approx \left[1+\Omega^4 (2\tau^2/n)^2 \delta^2 
(N-1)/2 \right]^{-2}$. 
Hence, the fidelity is improved, $F_{\rm sym}^n>F_{\rm st}$, if 
$n>2\Omega \tau$.

\begin{figure}
\includegraphics[width=6.cm,clip=true]{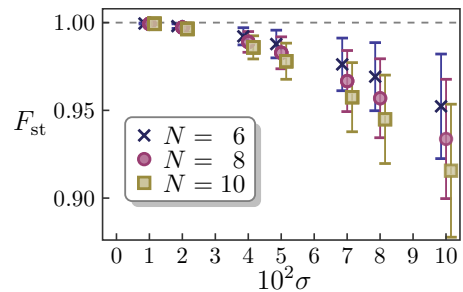}
\caption{Dependence of the cluster-state fidelity for
  $\tau=\pi/(4\Omega)$ on random errors in
the pulse duration for one-dimensional $XY$ qubit arrays of length
$N=6$, $8$, and $10$.  The width of the Gaussian distribution is
denoted by $\sigma$; the error bars indicate the standard deviation.}
\label{Fig2}
\end{figure}

These perturbative results are complemented by exact numerical
calculations of $F_{\mathrm{st}}(\tau)$ and the gate fidelity
$F_{\mathrm{g}}(\tau)=
2^{-N}\:|\mathrm{Tr}\:U_\tau^{\dagger}(0)U_\tau(\{\delta_i\})|$ for
$\tau=\pi/(4\Omega)$ in systems with up to $10$ qubits.  Here,
$\delta_i$, $i=1,\ldots,N-1$ corresponds to the qubit pair $(i,i+1)$.
In both the Ising and $XY$ case, we averaged these fidelities over
2000 random realizations of the $\delta_i$ taken from a Gaussian
distribution with varying width $\sigma$.  The results indicate that
the method is rather robust even outside the regime where $\sigma$ 
is much smaller than $\pi/4\approx 0.78$. For instance, the
$XY$-model calculation shows that $F_{\mathrm{st}}$ is bigger than
$99\%$ for $\sigma \lesssim 0.04$, while $F_g$ can be bigger than
$99\%$ even for $\sigma$ as large as 0.07. The comparison between the
two models shows that the robustness in the $XY$ case (see
Fig.~\ref{Fig2}) is somewhat better than in the Ising case. 

\section{Discussion and conclusion}
 
The distinguishing feature of the stabilizer Hamiltonians discussed
here is that their ground states are directly related to the universal
resource of measurement-based quantum computation.  More generally,
the stabilizer formalism has further applications in quantum
information theory \cite{NielsenChuang}.  For instance, in quantum
error-correcting codes~\cite{Gottesman}, the stabilizer
formalism is used to express codewords. The method illustrated in
Eq.~(\ref{rho1}) can be used to obtain eigenvalues for the syndrome
measurements in the process of detecting errors.  Moreover, relations
(\ref{Ising2}) and (\ref{XY3}) can be used to effectively generate
codewords in solid-state qubits.  For example, 
the three-qubit Greenberger-Horne-Zeilinger (GHZ) 
state $(|000\rangle \pm |111\rangle)/\sqrt{2}$, which is used for the
nine-qubit code, is effective generated by 
$e^{\pm i(\pi/4) \sigma_1^x
  \sigma_2^y\color{black} \sigma_3^x}|000\rangle$, where $e^{i(\pi/4) \sigma_1^x
  \sigma_2^y\color{black} \sigma_3^x}= e^{i(\pi/2) (\sigma_1^y+\sigma_3^y)} e^{\pm
  i(\pi/4) \sigma_1^z \sigma_2^y\color{black} \sigma_3^z} e^{-i(\pi/2)
  (\sigma_1^y+\sigma_3^y)}$ is applied to the three-qubit array.  The
five-qubit code \cite{Gottesman} can also be generated by using this
method.

To conclude, we have shown that initially prepared cluster states,
e.g., 2D cluster states that are universal resources of
measurement-based quantum computation, can be preserved with high
fidelity.  This is achieved by inducing the effective dynamics of 2D
stabilizer Hamiltonians by means of specially tailored pulse
sequences, starting from natural qubit-qubit interactions.  We have
also shown how this procedure can be implemented in the case of
always-on interactions. Our work will facilitate implementations of
one-way quantum computing.

\acknowledgements
We would like to thank J. Koga and F. Nori for discussions.  This work
was financially supported by the EU project SOLID, the Swiss SNF, the
NCCR Nanoscience, and the NCCR Quantum Science and Technology.

\end{document}